\documentclass{article}
\usepackage{iclr2025_conference,times}

\usepackage{amsmath,amsfonts,bm}

\def\eqref#1{equation~\ref{#1}}

\def\1{\bm{1}}

\DeclareMathAlphabet{\mathsfit}{\encodingdefault}{\sfdefault}{m}{sl}
\SetMathAlphabet{\mathsfit}{bold}{\encodingdefault}{\sfdefault}{bx}{n}

\usepackage{xreview}
\usepackage{tabularx}
\usepackage{multirow}
\usepackage{float}
\usepackage{hyperref}
\usepackage{url}
\usepackage{xspace}
\usepackage{xcolor}
\usepackage{booktabs}
\usepackage{graphicx}
\usepackage{placeins}

\title{\large OpenHarmony Bench: Evaluating LLMs and Coding Agents on OpenHarmony App Development}

\author{\makebox[\linewidth][c]{AI-Assisted R\&D TMG, Huawei}\\
\makebox[\linewidth][c]{SMAT Lab, Beihang University}\\
\makebox[\linewidth][c]{CodeWisdom Lab, Fudan University}}

\newcommand{\sysa}{\mbox{\textsc{OpenHarmony Bench}}\xspace}

\iclrfinalcopy 
\begin{document}
\raggedbottom

\maketitle
\begin{center}
\small\textbf{Official website:} \url{https://bench.matrix.openharmony.cn/}
\end{center}
\lhead{} 

\begin{abstract}
We present \textbf{OPENHARMONY BENCH}, an app-level coding benchmark for
evaluating LLM-based coding agents on OpenHarmony ArkTS applications. Unlike
function-level code generation benchmarks, \sysa evaluates complete app-level
changes: each task requires an agent to modify a buildable ArkTS project so
that a requested application behavior works end to end, which in turn requires
coordinated edits across project structure, UI state, local data persistence,
build configuration, and platform APIs. The benchmark then installs and drives
the delivered application on a device to check whether that behavior is actually
observable.
The benchmark covers three input sources: natural-language feature requests
(\texttt{new-feature}), structured scenario specifications
(\texttt{spec-driven}), and bug descriptions (\texttt{bug-fix}). The benchmark
contains 153 top-level inputs and 242 Feature points (F-points), where an
F-point is one executable behavior check. The current snapshot contains 32
new-feature tasks, 50 spec-driven tasks with 139 F-points, and 71 bug-fix tasks.
The main leaderboard is scored over top-level tasks rather than by independently
weighting F-points. We describe the benchmark construction, statistics, and
build-and-test evaluation pipeline, and evaluate DevEco Code with eight LLMs
across three independent full-suite runs per configuration, reporting the
arithmetic mean of top-level Task Completion across runs. Three findings emerge.
First, within the evaluated model-family pairs, newer generations complete more
tasks than their predecessors. Second, buildability is close to saturated while
behavioral correctness is not: mean Final Build Success Rate lies between
94.77\% and 100.00\%, whereas mean Task Completion lies between 48.36\% and
58.39\%, so a large fraction of delivered projects compile without implementing
the requested behavior. Third, spec-driven tasks have the lowest Task
Completion under all-checks task scoring, with no configuration exceeding
35\%.
The code, data, tasks, reference solutions, tests, evaluation scripts, and
leaderboard are released through the official \sysa website at
\url{https://bench.matrix.openharmony.cn/}.
\end{abstract}

\section{Introduction}
\label{sec:intro}

Large language models (LLMs) and coding agents are increasingly expected to
operate inside real software projects rather than solve isolated programming
exercises. Existing benchmarks have made substantial progress on function-level
code generation~\citep{chen2021evaluating,austin2021program,hendrycks2021apps,lu2021codexglue},
repository-level software maintenance~\citep{jimenez2024swebench,yang2024sweagent,xia2025agentless,multiswebench2025,xu2025swecompass},
and visual or web-oriented development tasks~\citep{yang2024swebenchmultimodal,cui2024webapp1k,xiao2025designbench}.
However, these settings do not fully capture application-level delivery.
Many coding benchmarks evaluate a patch with function-level, issue-level, or
unit-test-style checks, while mobile and OpenHarmony application tasks require
the submitted project to build, launch, and exhibit the expected GUI behavior in
a platform runtime. In this setting, correctness depends on ArkUI components,
page routing, UI state, persistence, platform APIs, resources, and build
configuration working together inside a runnable application. A benchmark for
OpenHarmony app development therefore needs to test whether a coding agent can
turn source-level changes into buildable and behaviorally correct applications,
rather than only satisfy narrower localized checks.

OpenHarmony is a useful app-development ecosystem for this style of evaluation.
Its ArkTS and ArkUI ecosystem~\citep{huawei_arkui_docs} is widely used in
HarmonyOS application development, but it poses several practical challenges
for benchmark construction. Tasks must be evaluated inside buildable
applications rather than standalone files. UI behavior must be checked through
emulators or devices. Platform APIs, project layouts, and build tools evolve
quickly, so benchmark tasks can become stale for reasons unrelated to agent ability.
In addition, many failures only appear after the application is built,
installed, launched, and tested through realistic GUI interactions. These
properties make OpenHarmony app development a concrete setting for evaluating
whether coding agents can deliver platform-specific application behavior under
realistic engineering requirements.

We present \textbf{OPENHARMONY BENCH}, an app-level coding benchmark for
OpenHarmony development. The benchmark is organized around three related input
sources that coding agents commonly face in OpenHarmony app development:
natural-language feature requests, structured scenario specifications, and bug
descriptions. The new-feature source contains 32 tasks in existing projects,
where each task starts from a concise natural-language description of a feature
to implement. The spec-driven source contains 50 structured tasks and 139
F-points, where each task starts from a scenario-level specification that may
cover multiple observable behaviors. The bug-fix source contains 71 real
bug-fix tasks, where each task starts from a description of an observed
application failure. We also characterize the platform
capabilities exercised by the benchmark through aggregate Kit/API coverage
analysis over all 153 tasks.

\sysa evaluates submitted projects through a shared build-and-test pipeline.
For each task, an agent produces or modifies an OpenHarmony project, the
pipeline checks buildability, and task-specific executable UI tests validate
observable behavior. Task Completion is the primary
effectiveness metric: one top-level benchmark input is marked complete only
when the final project builds and all required executable checks pass. The
spec-driven source still contains F-points as fine-grained checks inside each
task, and all F-point checks must pass for the corresponding
spec-driven task to be marked complete. Build pass rates and raw resource-use diagnostics provide
complementary signals. This separation is important because an implementation may build
successfully while still failing required interaction logic, state transitions,
or persistence behavior.

This report describes the benchmark construction, statistics, evaluation
pipeline, and benchmark-level baseline results. Each Agent--LLM configuration is
evaluated through three independent full-suite runs. We first compute Task
Completion for each run and then report the arithmetic mean across runs. The
leading configuration, DevEco Code with GLM-5.2, achieves a mean Task
Completion of 58.39\%, with an observed run-level range of
[56.21\%, 60.78\%]. Mean Final Build Success Rate remains high across the
evaluated configurations, ranging from 94.77\% to 100.00\%, while mean Task
Completion ranges from 48.36\% to 58.39\%. This gap indicates that
behavior-level correctness, rather than compilation alone, remains the main
challenge for OpenHarmony app-level coding agents.

\begin{figure}[!t]
\centering
\scriptsize
\begin{tikzpicture}[x=0.075cm,y=0.68cm]
  \draw[->,gray!65] (20,-0.65) -- (82,-0.65);
  \foreach \x in {20,40,60,80} {
    \draw[gray!25] (\x,-0.65) -- (\x,3.55);
    \node[anchor=north,gray!70] at (\x,-0.82) {\x};
  }
  \node[anchor=north] at (51,-1.28) {Task Completion (\%)};

  \node[anchor=east,font=\bfseries] at (-2.6,3.2)
    {Overall~{\scriptsize\textit{(GLM-5.2)}}};
  \draw[gray!25,line width=1.1pt] (20,3.2) -- (80,3.2);
  \fill[blue!30] (20,2.92) rectangle (58.39,3.48);
  \draw[blue!65!black,thick] (20,2.92) rectangle (58.39,3.48);
  \draw[black,thick] (56.21,3.2) -- (60.78,3.2);
  \draw[black,thick] (56.21,3.05) -- (56.21,3.35);
  \draw[black,thick] (60.78,3.05) -- (60.78,3.35);
  \filldraw[blue!70!black,draw=white,line width=0.4pt] (58.39,3.2) circle[radius=1.6pt];
  \node[anchor=west,font=\bfseries] at (81.2,3.2) {58.39};

  \draw[gray!35] (-17,2.55) -- (82,2.55);

  \foreach \y/\meanv/\minv/\maxv/\barname/\modelname in {
    2/63.54/59.38/65.62/New-feature/GLM-5.2,
    1/34.67/28.00/40.00/Spec-driven/Kimi-K3,
    0/75.12/70.42/78.87/Bug-fix/Kimi-K3
  } {
    \node[anchor=east] at (-2.6,\y)
      {\barname~{\scriptsize\textit{(\modelname)}}};
    \draw[gray!20,line width=1.1pt] (20,\y) -- (80,\y);
    \fill[blue!18] (20,\y-0.23) rectangle (\meanv,\y+0.23);
    \draw[blue!55!black,thick] (20,\y-0.23) rectangle (\meanv,\y+0.23);
    \draw[black,thick] (\minv,\y) -- (\maxv,\y);
    \draw[black,thick] (\minv,\y-0.12) -- (\minv,\y+0.12);
    \draw[black,thick] (\maxv,\y-0.12) -- (\maxv,\y+0.12);
    \filldraw[blue!65!black,draw=white,line width=0.4pt] (\meanv,\y) circle[radius=1.5pt];
    \node[anchor=west] at (81.2,\y) {\meanv};
  }
\end{tikzpicture}
\caption{Leading Task Completion by benchmark scope. Bars show the best mean
Task Completion across evaluated DevEco Code--LLM configurations for each
scope, labels identify the corresponding leading model, and whiskers show the
observed run-level min--max range for the same configuration. The x-axis starts
at 20\% to focus on the observed score range.}
\label{fig:task-completion-range-summary}
\end{figure}
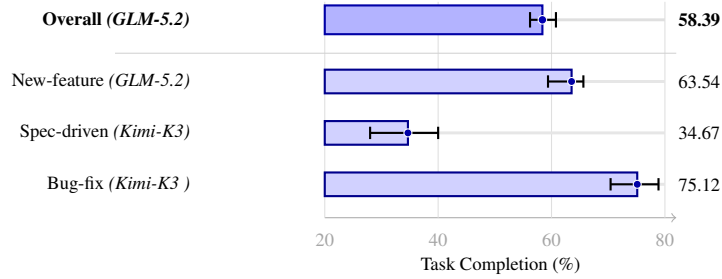

Figure~\ref{fig:task-completion-range-summary} summarizes the current
leading Task Completion result over the complete benchmark and the three task
sources introduced above. Here ``Overall'' aggregates all 153 tasks across
new-feature, spec-driven, and bug-fix. The top row is the complete-benchmark
aggregate, while the lower rows report the three task sources separately. Each
bar reports the best mean Task Completion among the evaluated DevEco Code--LLM
configurations for that scope; the label names the corresponding leading model,
and the whisker shows that configuration's observed run-level min--max range.

The contributions of this report are:
\begin{itemize}
  \item \textbf{An OpenHarmony app-level benchmark.} We define a benchmark
  structure spanning requirement-style incremental development,
  specification-driven development, and bug fixing.
  \item \textbf{Executable evaluation for ArkTS applications.} We describe a
  build-and-test pipeline based on OpenHarmony project builds and executable UI
  behavior checks, with Task Completion as the primary metric and buildability
  as a complementary metric.
  \item \textbf{Benchmark-level agent and model findings.} Under the fixed
  DevEco Code agent framework, we report four benchmark-level observations:
  evaluated configurations complete roughly half of the 153 top-level tasks;
  newer generations complete more tasks than their predecessors within the
  evaluated model-family pairs; Final Build Success Rate is high while Task
  Completion remains much lower; and spec-driven tasks have the lowest Task
  Completion under all-checks scoring.
\end{itemize}

\paragraph{Availability, versioning, and reproducibility.}
The benchmark artifacts needed to inspect and reproduce the current evaluation
snapshot are released through the official \sysa website, including task
descriptions, source-code scaffolds or repository pointers, reference solutions,
executable tests, evaluation scripts, and result aggregation code. The official
project URL is \url{https://bench.matrix.openharmony.cn/}. The results in this
report correspond to the \texttt{OpenHarmony Bench v1.0} task snapshot dated
\texttt{2026.7.31}.
Future leaderboard entries should report the benchmark version, task snapshot,
evaluation script snapshot, DevEco Code version, model identifier, execution
environment, and run configuration so that scores remain comparable over time.

\section{Benchmark Construction}
\label{sec:construction}

\sysa is an OpenHarmony app-level coding benchmark covering repository-level
incremental development and bug fixing. The new-feature source captures requirement-style
incremental feature implementation, the spec-driven source captures standardized
specification-driven incremental development, and the bug-fix source captures bug
fixing from real bug-fix tasks. Together, these task sources move beyond isolated code
snippets by requiring agents to produce buildable ArkTS applications and
validate the resulting behavior through build and test pipelines.

\subsection{Task Definition}
\label{sec:construction-task}

We formulate \sysa as an app-level coding benchmark for OpenHarmony
applications. Each instance provides a task description paired with a runnable
ArkTS project at a base commit. Given this input, a coding agent must modify the
project to satisfy the requested application behavior while preserving a
buildable OpenHarmony project. The agent-facing objective is the stated user
requirement rather than the executable checks. The executable checks are not
exposed to the agent. They are used only after generation to determine whether
the delivered project satisfies the requested behavior. This separation prevents
agents from optimizing against the oracle and keeps the task close to how
development requirements are given in practice.
The benchmark therefore rewards behaviorally correct app-level implementations
rather than token-level similarity to a reference patch.

The benchmark uses two units. The scoring unit is one top-level benchmark
input: one new-feature requirement, one spec-driven specification, or one
bug-fix report. The 153 inputs are therefore 153 tasks, and the leaderboard is
computed over these top-level tasks regardless of task source. The checking
unit is the \emph{Feature point} (F-point), which is one executable behavior
check. Most tasks have exactly one F-point, so the scoring unit and checking
unit coincide. Spec-driven tasks are the exception: a single specification can
describe several observable behaviors, so it carries several F-points
(2.78 on average). Such a task is scored all-or-nothing: it is marked complete
only if the project builds and every one of its F-points passes. F-points
therefore provide finer-grained diagnostic evidence within a task, but they are
not an independent leaderboard unit.

Compared with function-level code generation benchmarks, OpenHarmony app tasks require reasoning over project structure, routing, state management, persistence, UI components, build configuration, and platform APIs across multiple files. The task boundary is intentionally practical: the agent is expected to implement a specified application behavior, extend an existing project, or fix a known bug, while open-ended product design and requirement discovery are out of scope.

\subsection{Benchmark Task Sources}
\label{sec:construction-sources}

\sysa organizes these task families into three benchmark task sources. Table~\ref{tab:benchmark-tracks} summarizes their roles and evaluation signals.

\begin{table}[t]
\centering
\small
\setlength{\tabcolsep}{5pt}
\begin{tabularx}{\linewidth}{llX}
\toprule
Task source & Task type & Source and evaluation signal \\
\midrule
new-feature & Requirement-style feature implementation & Realistic requirement-style tasks with executable UI tests. \\
spec-driven & Specification-driven implementation & Standardized scenario-level tasks with 139 F-points implemented as Hypium checks. \\
bug-fix & Bug fixing & PR-derived bug-fix tasks whose tests fail before the fix and pass after the fix. \\
\bottomrule
\end{tabularx}
\caption{Benchmark task sources in \sysa.}
\label{tab:benchmark-tracks}
\end{table}

\paragraph{new-feature.}
The new-feature source evaluates incremental feature implementation in
OpenHarmony applications. Its 32 tasks are built primarily from Huawei official
template projects and requirement-style development scenarios, covering new
business features and functional iterations. Each task is screened to have a
clear requirement, a bounded implementation scope, and executable UI tests. This
source measures whether a coding agent can understand an existing project,
modify the relevant files, and deliver a feature that works in the application
context.

\paragraph{spec-driven.}
The spec-driven source contains 50 structured tasks and 139 F-points.
The 50 specifications are derived from open-source OpenHarmony projects through
preprocessing, manual screening, and curator-guided merging of related
requirements.
Unlike requirement-style tasks, spec-driven tasks use standardized
scenario-level specifications as the model-facing input. Each task is paired
with a runnable base project and a curator-reviewed reference implementation
that defines the expected behavior. The agent must translate each task into a
buildable, runnable, and verifiable OpenHarmony application implementation.

\paragraph{bug-fix.}
The bug-fix source evaluates engineering-level bug fixing for OpenHarmony ArkTS
projects in a repository-level setting exemplified by SWE-bench~\citep{jimenez2024swebench}.
Its 71 tasks are derived from bug-fix pull requests in real open-source
OpenHarmony projects. Candidate fixes are retained only when the bug is
reproducible, the fix scope is bounded, and the associated tests can distinguish
the buggy and fixed versions. This source targets fault localization, code
understanding, and bug-fixing effectiveness rather than new feature
construction.

\subsection{Task Construction}
\label{sec:task-construction}

All task sources are built around runnable OpenHarmony applications rather than
isolated code snippets. For each candidate input, we identify a base project
state, define the expected behavior, and attach task-specific automated checks.
Candidate tasks are retained only when the requirement is clear, the behavior is
reproducible, the required change is bounded, and the project remains compatible
with the evaluation environment. For bug-fix, we additionally screen for focused
underlying changes so that the task emphasizes fault localization and repair
rather than broad refactoring.

The new-feature source is primarily built from Huawei official template projects\footnote{\href{https://developer.huawei.com/consumer/cn/market/landing/component}{https://developer.huawei.com/consumer/cn/market/landing/component}}
and covers common development scenarios such as cross-layer feature
implementation, standalone UI feature implementation, HarmonyOS Kit
integration, third-party SDK integration, feature-parameter extension, and
scenario coverage extension. The bug-fix source is built from Huawei official
open-source repositories by selecting historical pull requests that cover bug
fixes, feature restoration, and stability optimization.

The spec-driven source is built from manually curated, standardized
scenario-level specifications over curated OpenHarmony base projects. For each
specification, curators select a compatible base project, implement a reference
solution, and retain only tasks whose expected behavior can be exercised
reliably in the shared OpenHarmony evaluation environment.

The new-feature and spec-driven sources differ in the task input given to the
agent. The new-feature source preserves compact development requests that
resemble real incremental requirements, while spec-driven uses standardized
scenario-level specifications with explicit acceptance structure. This
separation lets the benchmark cover realistic requirement-style feature work and
fine-grained specification-following analysis. The bug-fix source extends the
same project-level setting to tasks grounded in real bug-fix changes.

\begin{figure}[t]
\centering
\fbox{\parbox{0.94\linewidth}{
\footnotesize
\textbf{new-feature: requirement-style feature implementation.}
Add a Text component at the first position below the text-type components on the home page. The home-page component list is data-driven, so the Text entry must be added to both the data source and the code configuration. The new component should follow the existing component pattern, such as \texttt{TextInput}, and include a preview area plus a control panel with font weight, font size, text shadow, and letter spacing controls. The preview defaults to ``Hello Developer''.

\medskip
\textbf{spec-driven: wx-combo-chat-input task.}
Scenario 1: tapping the ``+'' icon expands the bottom action tray with an animation and reveals options such as Album, Camera, Voice Call, and Location. Scenario 2: long text automatically expands the input box to multiple lines. Scenario 3: entering text shows the green ``Send" button, and tapping ``Send" appends the message to the chat list. Scenario 4: deleting all text hides ``Send" and restores the ``+'' icon.

\medskip
\textbf{bug-fix: Penkit crash task.}
When the application is launched and the user clicks the ``Try Penkit'' button, the application crashes immediately, preventing normal entry into the corresponding feature module. The agent must localize and fix the bug so that the Penkit feature opens normally.
}}
\caption{Representative task examples from the three benchmark task sources in \sysa.}
\label{fig:benchmark-task-examples}
\end{figure}

\subsection{F-point and Test Construction}
\label{sec:construction-fpoints}
For all three task sources, benchmark curators manually construct executable
oracles and Hypium test scripts from the task description, expected behavior,
and project context. Each test is written to check user-visible behavior rather
than implementation-specific code structure.

Spec-driven tasks require additional normalization because one document often
combines main paths, alternative flows, and boundary conditions. We normalize
each spec-driven specification into a hierarchy of specification, scenario, and
F-point. A specification defines the full task scope, a scenario represents a
coherent user path, and an F-point corresponds to one executable behavior check
that should be implemented and verified independently. This construction yields
139 F-points for the spec-driven source. A spec-driven task is marked complete
only when the project builds and all F-point checks under the corresponding
task pass.

Every retained F-point must satisfy a fail-to-pass criterion. For new-feature
and spec-driven tasks, the corresponding test must fail on the base revision
where the requested behavior is absent and pass on the reference solution. For
bug-fix tasks, the test must reproduce the reported failure on the buggy
revision and pass on the fixed revision. Tests that do not distinguish the
pre-change and post-change states are discarded. At least two curators
independently validate each retained test, and each F-point must pass on the
reference solution in at least two validation runs before the task is admitted
into the benchmark.

Figure~\ref{fig:benchmark-task-examples} illustrates how this construction maps
to concrete tasks. In the new-feature example, the task has one F-point: the
delivered application must expose the requested Text component with the expected
preview and controls. In the bug-fix example, the task also has one F-point:
the Penkit entry must no longer crash, and the target feature module must open
normally. The spec-driven example is different because the specification
contains four required behaviors. Curators therefore define four F-points for
the chat-input task: expanding the action tray from the ``+'' icon, expanding
the input box for long text, showing the Send button and appending the sent
message to the chat list, and restoring the ``+'' icon after the input is
cleared. The spec-driven task is counted as complete only if the project builds
and all four F-points pass.

\section{Benchmark Statistics}
\label{sec:statistics}

We analyze OPENHARMONY BENCH at two levels. First, we report benchmark-wide
composition and task-source scale using dimensions with a common denominator.
We then examine workload scale, source concentration, difficulty, and aggregate
Kit/API coverage for the benchmark as a whole.

\subsection{Benchmark Composition and Scale}
\label{sec:statistics-composition}

Table~\ref{tab:benchmark-statistics} provides a single statistical view of all
three task sources. The benchmark contains 153 top-level inputs and 242
F-points drawn from 42 unique repositories. The repository total is a
union: new-feature contributes 12 repositories, bug-fix contributes 31, these
two sources share 11 repositories, and spec-driven contributes ten additional
repositories, yielding $12 + 31 - 11 + 10 = 42$ unique repositories.

\begin{table}[!t]
\centering
\scriptsize
\setlength{\tabcolsep}{3.0pt}
\textbf{(a) Common benchmark composition}\\[2pt]
\begin{tabular}{@{}lp{0.23\linewidth}rrrrc@{}}
\toprule
Task source & Input form & Repos & Inputs & F-points & Mean F-points per input &
Difficulty (E/M/H) \\
\midrule
new-feature & Requirement-style request & 12 & 32 & 32 & 1.00 & 11/18/3 \\
\shortstack[l]{spec-driven} & Structured task & 10 & 50 & 139 & 2.78 & 11/25/14 \\
bug-fix & Bug report & 31 & 71 & 71 & 1.00 & 36/29/6 \\
\midrule
\textbf{Overall} & -- & \textbf{42 repos} & \textbf{153} & \textbf{242} & \textbf{1.58} & \textbf{58/72/23} \\
\bottomrule

\end{tabular}

\vspace{3pt}
\textbf{(b) Input, reference, and repository-concentration statistics}\\[2pt]
\setlength{\tabcolsep}{2.2pt}
\begin{tabular}{@{}lrrrrr@{}}
\toprule
Task source & Input length [IQR] & Ref. files & Ref. LOC &
Largest repo (\%) & Top-3 repos (\%) \\
\midrule
new-feature & 118.5 [61.8--165.2] & 4 & 88 & 31.2\% & 53.1\% \\
\shortstack[l]{spec-driven} & 589 [496.5--794.5] & 12 & 489.5 & 38.0\% & 74.0\% \\
bug-fix & 107 [40--248.5] & 2 & 83 & 14.1\% & 31.0\% \\
\bottomrule

\end{tabular}

\caption{Benchmark-wide and task-source-level statistics. A repository source is one
OpenHarmony code repository from which benchmark inputs are drawn. Mean F-points
per input is the total number of executable checks divided by top-level inputs,
and difficulty reports easy/medium/hard. Input
length reports median [IQR] in non-whitespace characters. Ref. files and Ref.
LOC are medians of changed files and added-plus-deleted lines in the reference
patch. Largest repo and Top-3 repos report the percentage of inputs
coming from the most frequent repository and the three most frequent
repositories; percentages are used because the three task sources contain
different numbers of inputs. Kit/API coverage is reported separately at the
benchmark level in Section~\ref{sec:statistics-kit-coverage}.}
\label{tab:benchmark-statistics}
\end{table}

The spec-driven source has denser evaluation, with 2.78 F-points per top-level
task on average, while each new-feature requirement and bug-fix bug report has
one F-point. This difference reflects verification granularity rather than a
change in the leaderboard unit: the leaderboard is scored over top-level tasks,
and F-points are independently verifiable behavior checks within those tasks.

Input size shows the same distinction. The median spec-driven task contains
589 non-whitespace characters, compared with 119 for new-feature and 107 for
bug-fix. These values measure how much behavior is stated to the agent.
They are not interpreted as direct difficulty scores. The median reference
patches change four files and 88 lines for new-feature, 12 files and 489.5
lines for spec-driven, and two files and 83 lines for bug-fix. The spec-driven
values are computed over the 50 reference patches associated with the
spec-driven inputs. The 50 spec-driven inputs are distinct task
specifications with separate task files and Hypium checks. When a single app
state contains several independently verifiable behavior groups, curators may
split the original specification into multiple smaller tasks to keep each task
bounded and interpretable. Such tasks can share the same base and reference
state, but they remain separate evaluation inputs because they expose different
behavior requirements and different executable checks. This footprint describes
the reference implementation associated with an evaluation input rather than an
independent edit size for each F-point. Patch size also does not capture the
repository exploration needed to diagnose a bug.

Repository concentration reveals how broadly each task source samples the
OpenHarmony ecosystem rather than over-representing a small number of projects.
The new-feature source draws 31.2\% of its inputs from its largest repository and
53.1\% from its three largest repositories. The corresponding spec-driven
shares are 38.0\% and 74.0\%. The bug-fix source is less concentrated, at
14.1\% and 31.0\%, and therefore provides the broadest repository spread among
the three task sources.

Across the complete benchmark, 58 inputs (37.9\%) are labeled easy, 72
(47.1\%) medium, and 23 (15.0\%) hard. Difficulty is descriptive metadata
assigned during curation from reference-patch size, changed-file count,
cross-Kit/API involvement, and curator judgment of the development scenario.
It summarizes benchmark composition and is not used as a calibrated
performance predictor.

\subsection{Kit/API Coverage}
\label{sec:statistics-kit-coverage}

Kit/API coverage characterizes the OpenHarmony platform surface represented by
the benchmark. We analyze all instances against 39 high-frequency Kits commonly
used in OpenHarmony application development. Their names follow the Huawei
Developer taxonomy \citep{huawei_harmonyos_docs}.

A Kit is counted for an input when the gold patch imports an \texttt{@kit.*}
module or a recognized legacy \texttt{@ohos.*} module, or when it references a
Kit-specific API class or method such as \texttt{PixelMap}, \texttt{AVPlayer},
or \texttt{cameraManager}. Tests, generated output, dependencies,
documentation, and resources are excluded. The analysis is performed over the
153 task-level golden patches used by the benchmark.

\begin{figure}[t]
\centering
\includegraphics[width=0.90\linewidth]{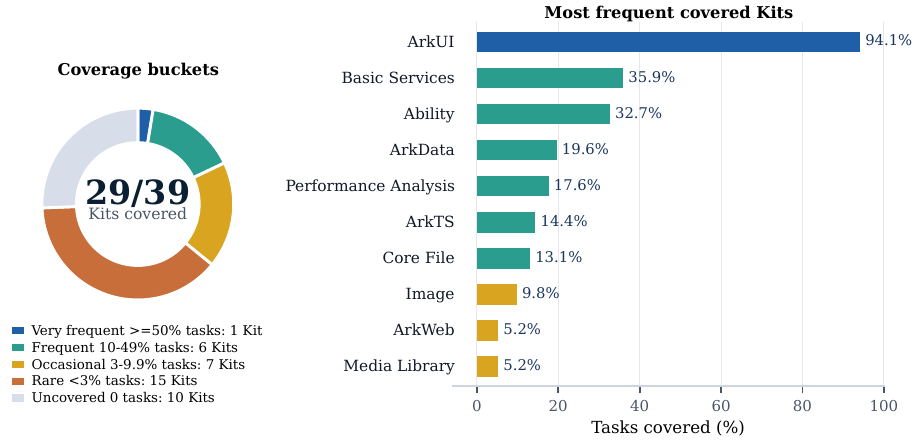}
\caption{Coverage of 39 high-frequency OpenHarmony Kits. The donut groups Kits
by task-level coverage frequency, i.e., the percentage of the 153 golden patches
in which each Kit appears. The bar chart shows the ten most frequent covered
Kits. Coverage describes the platform surface represented by the workload, not
model performance.}
\label{fig:kit-api-coverage-donut}
\end{figure}

Overall, the benchmark covers 29 out of 39 high-frequency Kits (74.4\%).
Figure~\ref{fig:kit-api-coverage-donut} explains this coverage in two ways. The
number at the center of the left donut, ``29/39 Kits covered'', reports the
number of high-frequency Kits touched by at least one gold patch. In the donut,
Kits are grouped by their task-level coverage frequency. A Kit is placed into
exactly one bucket according to
the percentage of the 153 benchmark tasks whose gold patch touches that Kit.
``Very frequent'' means the Kit appears in at least 50\% of tasks, ``Frequent''
means 10--49\%, ``Occasional'' means 3--9.9\%, ``Rare'' means greater than
0\% but less than 3\%, and ``Uncovered'' means the Kit appears in zero tasks.
Under this definition, one Kit is very frequent, six are frequent, seven are
occasional, fifteen are rare, and ten are uncovered. This shows that the
benchmark has broad Kit coverage but that most covered Kits appear in relatively
few tasks.

The right bar chart lists the ten covered Kits with the highest task-level
coverage frequency. Its horizontal axis, ``Tasks covered (\%)'', is the
percentage of the 153 tasks whose gold patch touches the corresponding Kit.
ArkUI appears in 144 of 153 tasks (94.1\%), reflecting the fact that most
patches involve ArkUI declarative UI components or DSL constructs such as
\texttt{struct}, \texttt{build()}, \texttt{\$r()}, and \texttt{@Component}.
Basic Services Kit and Ability Kit are the next most common platform surfaces,
appearing in 35.9\% and 32.7\% of tasks, followed by ArkData (19.6\%),
Performance Analysis Kit (17.6\%), ArkTS (14.4\%), and Core File Kit (13.1\%).
Image Kit, ArkWeb, and Media Library Kit complete the top-ten list but appear in
fewer than 10\% of tasks. A single task may exercise multiple Kits, so these
percentages describe workload composition rather than mutually exclusive
categories or model performance.

\section{Evaluation Pipeline}
\label{sec:setup}

\subsection{Pipeline Overview}

\sysa evaluates coding agents as end-to-end app developers. For each benchmark
instance, the evaluation pipeline prepares a clean workspace, provides the task
input to the agent, records the submitted code changes, builds the resulting
OpenHarmony project, and then runs executable checks whenever the project is
buildable. Every scored benchmark input has task-specific executable checks;
static and code-quality diagnostics are separate from the leaderboard score.
This pipeline is shared across incremental feature implementation,
specification-driven development, and bug fixing. The
task input differs by source, but the delivered artifact is always a buildable
ArkTS application project.

The pipeline has six stages:
\begin{enumerate}
  \item \textbf{Workspace initialization.} The pipeline initializes a clean
  workspace from the task scaffold or base project.
  \item \textbf{Agent execution.} The agent receives the task description,
  relevant project context, system instruction, prompt template, and tool
  access, then produces a modified workspace.
  \item \textbf{Diff recording.} The pipeline records whether the run produced a
  non-empty diff. This diagnostic separates no-output failures from later build
  or behavior failures, but it is not used as a correctness signal.
  \item \textbf{First build.} The evaluation protocol records the first
  project build before any compile-repair step.
  \item \textbf{Compile repair.} If the first build fails, the agent may enter a
  bounded compile-repair loop. The final submitted workspace is then built
  again.
  \item \textbf{Behavior validation.} If the final build succeeds, the dynamic
  evaluation scripts install the application, launch it as required, and run the corresponding
  Python Hypium checks.
\end{enumerate}

\subsection{Evaluation Dimensions}
\label{sec:evaluation-dimensions}

No single evaluation dimension can adequately assess app-level code generation.
A UI-only pass may miss incorrect persistence or platform integration, whereas
a static-only pass cannot confirm whether user-visible behavior actually works.
\sysa therefore records complementary dimensions: buildability, dynamic
behavior verification, and resource use. The current leaderboard is ranked by
top-level Task Completion. Buildability and resource use are reported as
diagnostics, while static functional coverage and code-quality checks are used
during benchmark construction or retained as optional diagnostics when
available.

Buildability checks whether the delivered ArkTS project can be compiled into a
runnable OpenHarmony application. Dynamic behavior verification executes
scenario-specific UI tests on OpenHarmony emulators or devices. Resource metrics
record the token use and agent wall-clock time needed to produce the submitted
workspace. Static functional coverage and code-quality checks are construction
and diagnostic signals rather than leaderboard metrics. Only the build and
dynamic behavior dimensions determine the current Task Completion scores.

\subsection{Execution Environment}

All experiments are executed with DevEco 6.1 and API 23 on an OpenHarmony
emulator configured with the Pura 90 Pro Max device profile, a resolution of
1308$\times$2880, and a 6.9-inch screen.
For compilation, we use the OpenHarmony API 23 SDK. The build pipeline is based on \texttt{hvigorw}~\cite{huawei_hvigor_docs}.
Dynamic behavior verification uses OpenHarmony's Hypium UI testing
framework~\cite{huawei_uitest_docs}. The dynamic evaluation scripts install application packages
with \texttt{hdc}~\cite{openharmony_hdc_repo}, launch the application when
needed, and invoke the Python Hypium runner for the task-specific UI checks.

When compile repair is enabled, we allow at most five compile-and-fix
iterations per task. A repair iteration receives only build output, such as
compiler diagnostics and build logs. Hypium outputs, oracle results, reference
patches, and task-specific expected-check details are not returned to the agent
during repair. Tokens and agent wall-clock time consumed by these repair
iterations are included in the resource metrics reported in
Table~\ref{tab:cross-benchmark-results}. The evaluated snapshot does not impose
an additional benchmark-level token or wall-clock cap beyond the shared
five-iteration repair limit and provider/tool constraints.
If compilation still fails after the allowed repair budget, the case is marked
as a build failure and receives zero dynamic passes. This separation lets us
distinguish generation failures, build failures, and behavioral failures after
a successful build.

\subsection{Agents and Models}
\label{sec:agents-models}

Unless otherwise stated, all reported results use DevEco
Code~\cite{deveco_code_docs} as the fixed agent framework. The main comparison
varies the underlying model, identified by the display names configured in
DevEco Code: GLM-5.1 and GLM-5.2~\citep{zhipu_glm51_docs,zhipu_glm52_docs},
Qwen3.7-Max and Qwen3.8-Max-Preview~\citep{alibaba_modelstudio_models,alibaba_modelstudio_home},
Kimi-K2.7-Code and Kimi-K3~\citep{kimi_k27_code,kimi_code_models},
MiniMax-M3~\citep{minimax_m3_model}, and
DeepSeek-V4-Pro~\citep{alibaba_modelstudio_models,deepseek_transparency}. This
organization focuses the current report on model differences under one shared
OpenHarmony coding-agent scaffold.

All runs for a given comparison use the same task split, system instruction,
prompt template, tool access, environment configuration, build scripts, and
scoring scripts. This keeps the DevEco Code agent framework fixed and isolates
the effect of the underlying model as much as possible. The main result table
reports the concrete model cells evaluated over the complete 153-task suite.
Results should therefore be read as DevEco Code--model configurations rather
than as a comparison of multiple independent agent frameworks.

\subsection{Evaluation Snapshot and Reproducibility}
\label{sec:reproducibility}

Table~\ref{tab:evaluation-snapshot} summarizes the exact snapshot evaluated in
this report. We report these identifiers because leaderboard scores depend not
only on the model, but also on the task set, dynamic evaluation scripts, agent framework, SDK,
device profile, and run configuration.

\begin{table}[H]
\centering
\small
\begin{tabularx}{\linewidth}{@{}p{0.30\linewidth}X@{}}
\toprule
Item & Snapshot used in this report \\
\midrule
Benchmark version & \texttt{OpenHarmony Bench v1.0} \\
Task snapshot date & \texttt{2026.7.31} \\
Official project URL & \url{https://bench.matrix.openharmony.cn/} \\
Number of top-level tasks & 153 \\
Task sources & new-feature 32; spec-driven 50; bug-fix 71 \\
Executable F-points & 242 total; spec-driven 139 \\
DevEco Code version & \texttt{0.1.0-1.3} \\
Prompt/template & \texttt{(1) Task definition; (2) Spec content and environment; (3) Code generation instructions} \\
Run parameters & \texttt{Temperature=0, Context-window=200K, Reasoning-budget=65.5K} \\
OpenHarmony SDK & API 23 \\
DevEco Studio / toolchain & DevEco 6.1 \\
Device profile & Pura 90 Pro Max emulator, 1308$\times$2880, 6.9-inch \\
Run configuration & 3 independent full-suite runs; at most 5 compile-repair iterations per task \\
\bottomrule
\end{tabularx}
\caption{Evaluation snapshot used for the results in this report.}
\label{tab:evaluation-snapshot}
\end{table}

\paragraph{Open artifacts and contamination safeguards.}
The public release includes the task descriptions, benchmark metadata,
reference patches or solutions, executable tests, evaluation scripts, and
aggregation scripts. This open design allows developers to inspect task
quality, reproduce scores, and run additional Agent--LLM configurations. Public
release may introduce training-contamination risk over time, and the bug-fix
source may also carry pre-training exposure risk because it is derived from
historical public pull requests. We therefore treat the reported numbers as a
timestamped snapshot, record benchmark versions and evaluation script snapshots
for every leaderboard entry, and recommend that future evaluations use updated
task snapshots and trajectory audits to check for reward hacking or overfitting
to released tests.

\subsection{Metrics}
\label{sec:metrics}
\begingroup
\setlength{\abovedisplayskip}{4pt}
\setlength{\belowdisplayskip}{4pt}
\setlength{\abovedisplayshortskip}{2pt}
\setlength{\belowdisplayshortskip}{3pt}

We use top-level Task Completion as the primary effectiveness metric and
leaderboard score. One evaluation task is one benchmark input: a new-feature
request, a spec-driven task, or a bug-fix input. A task is marked complete
only when the final project builds and all executable checks associated with
that input pass. Buildability is therefore necessary but not sufficient for
completion: a project can compile successfully while still failing required UI
behavior, state transitions, persistence writes, or platform integration. The
spec-driven source still contains multiple F-point checks inside each
task; a spec-driven task is marked complete only if all F-point
checks for that task pass. This top-level definition is stricter than
counting individual F-point checks independently because a task is marked
complete only when all of its checked behaviors pass. It mirrors the all-tests
task success view used in repository-level benchmarks.

The metric has an explicit floor and ceiling in the released snapshot. A no-op
baseline that submits the base or buggy project unchanged completes 0 of the
153 top-level tasks, because every retained executable check is required to
fail before the reference change and pass after it. The reference solutions
define a 100\% ceiling: a task is admitted only when its reference solution
builds and all retained executable checks pass under curator validation.

For the current benchmark, the overall leaderboard aggregates 153 top-level
tasks: 32 new-feature tasks, 50 spec-driven tasks, and 71 bug-fix tasks. Each
DevEco Code--LLM configuration is evaluated with three independent full-suite
runs. We compute every rate separately for each run and then report the
arithmetic mean across the three runs. Tables and prose report rates as
percentages.

For a reported scope $S$ (either all 153 tasks or one task source), let
$N=|S|$ and $k=3$. For configuration $c$ and run $r$, let
$C_{c,r}$ be the number of completed tasks in $S$, and let
$B^{\mathrm{first}}_{c,r}$ and $B^{\mathrm{final}}_{c,r}$ be the numbers of
tasks whose first and final build attempts succeed.

\paragraph{Task Completion.}
Task Completion is the fraction of top-level tasks that build and pass all
required executable checks in a run:
\[
\mathrm{TC}_{c,r}(S)=\frac{C_{c,r}}{N}, \qquad
\overline{\mathrm{TC}}_{c}(S)=\frac{1}{k}\sum_{r=1}^{k}\mathrm{TC}_{c,r}(S).
\]
The main leaderboard reports $\overline{\mathrm{TC}}_{c}(S)$ with $S$ equal to
all 153 tasks. Source-level tables apply the same rule with $S$ restricted to
one task source. This run-level averaging treats each complete-suite execution
as one repeated measurement of the same Agent--LLM configuration, and each run
contributes equally to the reported mean.

\paragraph{Run range.}
Because each configuration is evaluated three times, we report the observed
minimum and maximum run-level Task Completion beside the mean score:
\[
\mathrm{Range}_{c}(S)=
\left[
\min_{r\in\{1,\ldots,k\}}\mathrm{TC}_{c,r}(S),
\max_{r\in\{1,\ldots,k\}}\mathrm{TC}_{c,r}(S)
\right].
\]
This range is descriptive rather than an inferential interval, and no
best-of-three rule is applied. For the overall 153-task table, each full-suite
run first aggregates all task sources, and the minimum and maximum are then
taken across the three overall run scores. The overall range is not formed by
independently combining task-source-level extrema.


\paragraph{Final Build Success Rate.}
Final Build Success Rate measures whether the delivered workspace builds after
the complete agent run and any allowed compile-repair iterations. It is the
fraction of tasks whose submitted ArkTS project builds successfully at the end
of the agent run, averaged across the three runs:
\[
\mathrm{FBR}_{c,r}(S)=\frac{B^{\mathrm{final}}_{c,r}}{N}, \qquad
\overline{\mathrm{FBR}}_{c}(S)=\frac{1}{k}\sum_{r=1}^{k}\mathrm{FBR}_{c,r}(S).
\]
The Final Build column in Table~\ref{tab:cross-benchmark-results} reports this
metric over all 153 tasks.

\paragraph{First Build Success Rate.}
First Build Success Rate measures whether the project builds successfully on the
first build attempt, before any compile-repair iteration. The runner records
this first build outcome before source repair so that it can be distinguished
from the final build outcome:
\[
\mathrm{IBR}_{c,r}(S)=\frac{B^{\mathrm{first}}_{c,r}}{N}, \qquad
\overline{\mathrm{IBR}}_{c}(S)=\frac{1}{k}\sum_{r=1}^{k}\mathrm{IBR}_{c,r}(S).
\]
The First Build column reports the mean first build success rate across the
three runs.

\paragraph{Run-level resource metrics.}
We report token usage and agent wall-clock time as descriptive resource
dimensions for every Agent--LLM configuration. For each complete-suite run, we
first average the raw per-task resource values over all 153 inputs and then
average those per-run values across the three runs. For resource dimension $q$
(tokens or agent wall-clock time), let $x_{c,r,i}^{(q)}$ be the raw value for
task $i$ in run $r$:
\[
X_{c,r}^{(q)}=\frac{1}{N}\sum_{i\in S}x_{c,r,i}^{(q)}, \qquad
\overline{X}_{c}^{(q)}=\frac{1}{k}\sum_{r=1}^{k}X_{c,r}^{(q)}.
\]
For presentation, Table~\ref{tab:cross-benchmark-results} converts tokens to
millions per task and agent wall-clock time to seconds per task.
Token usage is the provider-recorded \texttt{total\_tokens} value accumulated
over every conversation turn. It includes cache-read or cache-creation tokens
when the provider reports them in \texttt{total\_tokens}; component fields are
retained in the raw records but are not added again. Agent wall-clock time is
measured from the start of the Agent process until it completes. It includes model
interactions, tool use, source modification, and any builds initiated by the
Agent. Post-run validation builds, installation, and Hypium execution are part
of correctness evaluation but are excluded from agent wall-clock time.

This time definition deliberately measures the execution time of the complete
Agent--LLM system, including local Agent actions.
Because agent wall-clock time includes local tool use, source modification, and
builds, it can vary with hardware performance, build-server load, filesystem
state, and emulator/device state. We therefore use wall-clock time as a
within-environment
diagnostic and do not treat it as directly comparable across different execution
environments.

These dimensions measure gross resource use rather than normalized efficiency.
A configuration that fails early can appear faster or cheaper because it does
less work. We therefore interpret wall-clock time and token use together with
Task Completion.
\endgroup


\section{Results}
\label{sec:results}

\begin{figure}[H]
\centering
\includegraphics[width=0.95\linewidth]{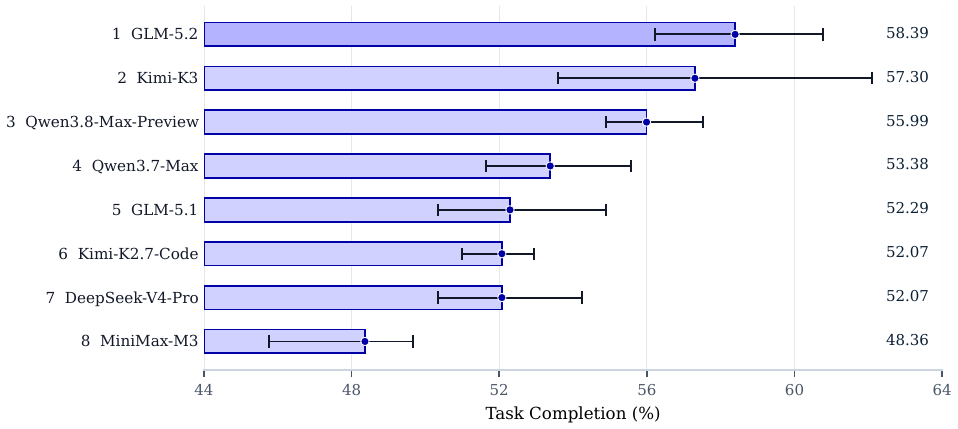}
\caption{Overall ranking over all 153 OpenHarmony tasks. Bars show mean Task
Completion across three independent full-suite runs, and whiskers show the
observed run-level min--max range. The x-axis starts at 44\% to focus on the
observed score range.}
\label{fig:overall-leaderboard-ranking}
\end{figure}

Figure~\ref{fig:overall-leaderboard-ranking} visualizes the main ranking, and
Table~\ref{tab:cross-benchmark-results} reports the full evaluation matrix.
Each row represents one DevEco Code--LLM configuration evaluated on all 153
tasks. Task Completion reports the arithmetic mean across three independent
full-suite runs, with the run-level minimum and maximum shown in brackets.
Build and resource columns are averaged over the same three runs.

\begin{table}[H]
\centering
\scriptsize
\setlength{\tabcolsep}{3.5pt}
\renewcommand{\arraystretch}{1.06}
\resizebox{\linewidth}{!}{%
\begin{tabular}{@{}rlccccc@{}}
\toprule
Rank & Model & \shortstack{Task Completion\\mean [min, max] (\%)} &
\shortstack{First Build\\(\%)} & \shortstack{Final Build\\(\%)} &
\shortstack{Tokens/Task\\(M)} & \shortstack{Time/Task\\(s)} \\
\midrule
1 & GLM-5.2 & \textbf{58.39 [56.21, 60.78]} & 67.10 & 98.04 & 2.57 & 1936.4 \\
2 & Kimi-K3 & 57.30 [53.59, 62.09] & \textbf{87.58} & 98.91 & 1.02 & 652.3 \\
3 & Qwen3.8-Max-Preview & 55.99 [54.90, 57.52] & 80.61 & 99.56 & \textbf{0.98} & \textbf{394.3} \\
4 & Qwen3.7-Max & 53.38 [51.63, 55.56] & 69.06 & 98.47 & 1.27 & 738.3 \\
5 & GLM-5.1 & 52.29 [50.33, 54.90] & 73.20 & 99.57 & 1.60 & 1097.8 \\
6 & Kimi-K2.7-Code & 52.07 [50.98, 52.94] & 62.53 & 99.35 & 1.88 & 767.5 \\
7 & DeepSeek-V4-Pro & 52.07 [50.33, 54.25] & 64.71 & \textbf{100.00} & 1.33 & 1048.6 \\
8 & MiniMax-M3 & 48.36 [45.75, 49.67] & 59.26 & 94.77 & 3.69 & 870.8 \\
\bottomrule
\end{tabular}
}
\caption{Main results over all 153 OpenHarmony tasks. Task Completion reports
the mean across three independent runs, with the run-level min--max range shown
in brackets. First Build and Final Build are agent-run build outcomes averaged
over the same runs. Tokens/Task uses provider-recorded \texttt{total\_tokens};
Time/Task reports mean per-task agent wall-clock time.}
\label{tab:cross-benchmark-results}
\end{table}
\subsection{Overall Performance}

Across three independent full-suite runs over all 153 tasks, DevEco Code with
GLM-5.2 achieves the highest mean Task Completion at 58.39\%
(approximately 89.3/153 tasks). Kimi-K3 ranks second at 57.30\%, and
Qwen3.8-Max-Preview ranks third at 55.99\%. These values are arithmetic means
of run-level rates and do not apply a best-of-three rule. The bracketed ranges
in Table~\ref{tab:cross-benchmark-results} show the minimum and maximum
run-level Task Completion values for each model.

The rank column should be read as a descriptive ordering for this three-run
snapshot rather than as a strict statistical separation between adjacent
configurations. The observed run range of Kimi-K3, [53.59, 62.09], contains the
mean of GLM-5.2, and Kimi-K2.7-Code and DeepSeek-V4-Pro differ only beyond the
two decimal places shown in the table. The table therefore supports a leading
group view: GLM-5.2, Kimi-K3, and Qwen3.8-Max-Preview form the top group under
the current DevEco Code setting, while adjacent fine-grained ranks should be
interpreted with caution.

\subsection{Model Performance}

Task Completion ranges from 48.36\% for MiniMax-M3 to 58.39\% for GLM-5.2.
The corresponding mean Final Build Success Rate ranges from 94.77\% for
MiniMax-M3 to 100.00\% for DeepSeek-V4-Pro. These results show that a high
final build rate does not by itself imply high behavioral correctness:
DeepSeek-V4-Pro achieves the highest Final Build Success Rate, while GLM-5.2
achieves the highest overall Task Completion.

Within the evaluated model families under the fixed DevEco Code setting, the
newer model variants obtain higher mean Task Completion than their earlier
counterparts. This observation is limited to the evaluated configurations and
does not by itself establish a general trend across all OpenHarmony coding
agents or model families.

The complete 153-task result combines the new-feature, spec-driven, and bug-fix
task sources using the same top-level Task Completion definition. We use this
overall task-level aggregation as the primary leaderboard unit. We also report
the three task sources separately to explain where model differences arise.

\subsection{Task-Source Results}

\begin{figure}[!htbp]
\centering
\includegraphics[width=0.98\linewidth]{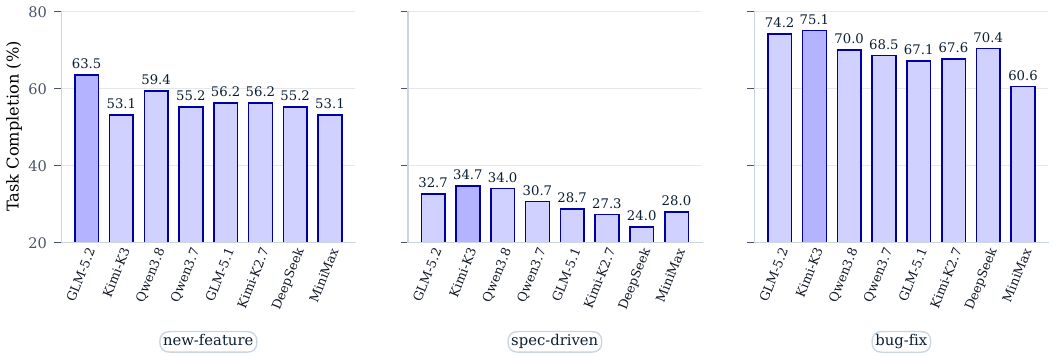}
\caption{Task Completion by task source for the evaluated DevEco Code--LLM
configurations. Each vertical bar represents one configuration. Each panel uses
the same model order and the same y-axis range;
bars report mean Task Completion across three independent full-suite runs. The
y-axis starts at 20\% to focus on the observed score range.}
\label{fig:source-leaderboard-panels}
\end{figure}

Figure~\ref{fig:source-leaderboard-panels} and
Tables~\ref{tab:new-feature-results}--\ref{tab:bug-fix-results} decompose the
153-task result into the three task sources. These source-level results are
diagnostic rather than separate benchmark leaderboards: the main ranking remains
the 153-task aggregation in Table~\ref{tab:cross-benchmark-results}. The
decomposition follows the report-wide task-source order: new-feature,
spec-driven, and bug-fix. In the current snapshot, bug-fix tasks have the
highest observed completion rates, new-feature tasks sit in the middle, and
spec-driven tasks have the lowest observed completion rates under the task-level
completion metric.

\begin{table}[!htbp]
\centering
\scriptsize
\setlength{\tabcolsep}{3.5pt}
\renewcommand{\arraystretch}{1.16}
\resizebox{\linewidth}{!}{%
\begin{tabular}{@{}rlccccc@{}}
\toprule
Rank & Model & \shortstack{Task Completion\\mean [min, max] (\%)} &
\shortstack{First Build\\(\%)} & \shortstack{Final Build\\(\%)} &
\shortstack{Tokens/Task\\(M)} & \shortstack{Time/Task\\(s)} \\
\midrule
1 & GLM-5.2 & \textbf{63.54 [59.38, 65.62]} & 60.42 & 96.88 & 2.08 & 1597.0 \\
2 & Qwen3.8-Max-Preview & 59.38 [59.38, 59.38] & 91.66 & 100.00 & 1.01 & 359.3 \\
3 & GLM-5.1 & 56.25 [53.12, 59.38] & 89.58 & 100.00 & 1.42 & 941.0 \\
4 & Kimi-K2.7-Code & 56.25 [53.12, 59.38] & 71.88 & 100.00 & 1.77 & 521.7 \\
5 & Qwen3.7-Max & 55.21 [46.88, 62.50] & 72.92 & 100.00 & 1.07 & 368.0 \\
6 & DeepSeek-V4-Pro & 55.21 [53.12, 56.25] & 61.46 & 100.00 & 1.00 & 328.3 \\
7 & MiniMax-M3 & 53.13 [43.75, 59.38] & 61.46 & 93.75 & 3.57 & 463.4 \\
8 & Kimi-K3 & 53.13 [43.75, 59.38] & 83.33 & 100.00 & 0.84 & 633.4 \\
\bottomrule
\end{tabular}
}
\caption{Results on the new-feature task source, covering 32 tasks.}
\label{tab:new-feature-results}
\end{table}

New-feature tasks require agents to implement requirement-style changes in
existing OpenHarmony applications. GLM-5.2 achieves the highest mean Task
Completion, but it also uses substantially more time than most other
configurations. Qwen3.8-Max-Preview ranks second while keeping both token use
and wall-clock time low. It completes 19 of 32 tasks in each of the three runs,
which is why its new-feature run range has the same minimum and maximum. The wider
min--max ranges for several models show that this
source is more sensitive to run-level variation than bug-fix.

\begin{table}[!htbp]
\centering
\scriptsize
\setlength{\tabcolsep}{3.5pt}
\renewcommand{\arraystretch}{1.16}
\resizebox{\linewidth}{!}{%
\begin{tabular}{@{}rlccccc@{}}
\toprule
Rank & Model & \shortstack{Task Completion\\mean [min, max] (\%)} &
\shortstack{First Build\\(\%)} & \shortstack{Final Build\\(\%)} &
\shortstack{Tokens/Task\\(M)} & \shortstack{Time/Task\\(s)} \\
\midrule
1 & Kimi-K3 & \textbf{34.67 [28.00, 40.00]} & 92.67 & 96.67 & 1.54 & 928.8 \\
2 & Qwen3.8-Max-Preview & 34.00 [32.00, 36.00] & 68.00 & 100.00 & 1.34 & 610.6 \\
3 & GLM-5.2 & 32.67 [30.00, 34.00] & 63.33 & 100.00 & 4.47 & 2971.6 \\
4 & Qwen3.7-Max & 30.67 [28.00, 32.00] & 62.67 & 100.00 & 2.06 & 1633.9 \\
5 & GLM-5.1 & 28.67 [28.00, 30.00] & 57.33 & 99.33 & 2.36 & 1857.9 \\
6 & MiniMax-M3 & 28.00 [28.00, 28.00] & 50.00 & 98.67 & 5.56 & 1746.7 \\
7 & Kimi-K2.7-Code & 27.33 [26.00, 28.00] & 53.33 & 100.00 & 2.86 & 1412.7 \\
8 & DeepSeek-V4-Pro & 24.00 [20.00, 26.00] & 61.33 & 100.00 & 2.17 & 2580.0 \\
\bottomrule
\end{tabular}
}
\caption{Results on the spec-driven task source, covering 50 tasks.}
\label{tab:spec-driven-results}
\end{table}

Spec-driven has the lowest observed completion under top-level completion
scoring. Its mean Task Completion ranges from 24.00\% to 34.67\%, lower than
the other two sources. This gap is consistent with the stricter task-level rule
for this source: a spec-driven task is marked complete only when all of its
required F-point checks pass. The source therefore stresses precise
specification following and end-to-end UI behavior, not just buildability:
several configurations reach a 100\% mean Final Build Success Rate while
remaining below 35\% Task Completion.

This source-level comparison should be read at the common top-level task unit.
Spec-driven tasks contain multiple F-points, whereas new-feature and bug-fix
tasks contain one executable check each. We therefore report source-level
results with the same task-completion rule used by the main leaderboard, rather
than adding a separate F-point-level ranking that would not align with the other
two task sources.

\begin{table}[!htbp]
\centering
\scriptsize
\setlength{\tabcolsep}{3.5pt}
\renewcommand{\arraystretch}{1.16}
\resizebox{\linewidth}{!}{%
\begin{tabular}{@{}rlccccc@{}}
\toprule
Rank & Model & \shortstack{Task Completion\\mean [min, max] (\%)} &
\shortstack{First Build\\(\%)} & \shortstack{Final Build\\(\%)} &
\shortstack{Tokens/Task\\(M)} & \shortstack{Time/Task\\(s)} \\
\midrule
1 & Kimi-K3 & \textbf{75.12 [70.42, 78.87]} & 85.91 & 100.00 & 0.74 & 466.1 \\
2 & GLM-5.2 & 74.18 [71.83, 77.46] & 72.77 & 97.18 & 1.46 & 1360.3 \\
3 & DeepSeek-V4-Pro & 70.42 [66.20, 73.24] & 68.54 & 100.00 & 0.89 & 294.8 \\
4 & Qwen3.8-Max-Preview & 69.95 [66.20, 73.24] & 84.51 & 99.06 & 0.71 & 257.6 \\
5 & Qwen3.7-Max & 68.54 [66.20, 70.42] & 71.83 & 96.71 & 0.80 & 274.5 \\
6 & Kimi-K2.7-Code & 67.61 [67.61, 67.61] & 64.79 & 98.59 & 1.24 & 423.9 \\
7 & GLM-5.1 & 67.14 [64.79, 71.83] & 77.00 & 99.53 & 1.14 & 633.2 \\
8 & MiniMax-M3 & 60.56 [59.15, 61.97] & 64.79 & 92.49 & 2.43 & 437.4 \\
\bottomrule
\end{tabular}
}
\caption{Results on the bug-fix task source, covering 71 tasks.}
\label{tab:bug-fix-results}
\end{table}

Bug-fix has the highest observed completion rates in the current benchmark
snapshot. Kimi-K3 and GLM-5.2 form the leading group, both exceeding 74\% mean
Task Completion.
DeepSeek-V4-Pro and Qwen3.8-Max-Preview follow closely, with Qwen3.8-Max-Preview
using the lowest mean tokens and wall-clock time on this source. The high final build
rates indicate that remaining failures are mostly behavioral rather than
compilation failures.

\subsection{Buildability and Functional Correctness}

Final Build and Task Completion measure different outcomes. Across the evaluated
DevEco Code configurations, mean Final Build Success Rate ranges from 94.77\% to
100.00\%, while mean Task Completion ranges from 48.36\% to 58.39\%. The gap
indicates that buildability is necessary but not sufficient: many submitted
projects compile after repair but still fail to satisfy the requested
behavior-level checks.

The highest mean First Build Success Rate is 87.58\% for Kimi-K3, while the
lowest is 59.26\% for MiniMax-M3. For the leading overall configuration,
GLM-5.2, the mean First Build Success Rate is 67.10\%, mean Final Build Success
Rate is 98.04\%, and mean Task Completion is 58.39\%. These values indicate how
often compile repair recovers first-build failures and how often a buildable
result also implements the required interaction, state transition, and UI
behavior.

\subsection{Resource Use and Run-to-Run Variation}

Mean per-task resource use varies across DevEco Code configurations. The lowest
mean token use is 0.98 million tokens per task for Qwen3.8-Max-Preview, while
the highest is 3.69 million tokens per task for MiniMax-M3. The lowest mean
agent wall-clock time is 394.3 seconds per task for Qwen3.8-Max-Preview, while
the highest is 1936.4 seconds per task for GLM-5.2. The leading effectiveness
configuration, GLM-5.2, uses a mean of 2.57 million tokens and 1936.4 seconds
per task. Token counts are not directly comparable across models, since each
provider may use a different tokenizer and accounting convention for cached
prefixes; the column indicates order of magnitude rather than a like-for-like
ratio. Raw resource totals must be interpreted together with effectiveness
because a configuration that stops early can consume fewer resources by
completing less work.

The bracketed ranges in Table~\ref{tab:cross-benchmark-results} summarize
observed run-to-run variability across the three full-suite runs. We report this
descriptive min--max range instead of an inferential interval because the current
evaluation uses three repeated runs per configuration.

\section{Related Work}
\label{sec:related}

\paragraph{Code and repository-level benchmarks.}
Early code benchmarks evaluate self-contained synthesis from docstrings,
signatures, or programming-problem statements, including
HumanEval~\citep{chen2021evaluating}, MBPP~\citep{austin2021program}, and
APPS~\citep{hendrycks2021apps}. CodeXGLUE~\citep{lu2021codexglue} broadens the
scope to multiple code-understanding and generation tasks, while
DS-1000~\citep{lai2023ds1000} and BigCodeBench~\citep{zhuo2024bigcodebench}
stress library use and complex instructions. Repository-level benchmarks such
as SWE-bench~\citep{jimenez2024swebench}, SWE-agent~\citep{yang2024sweagent},
Agentless~\citep{xia2025agentless}, SWE-bench Multimodal~\citep{yang2024swebenchmultimodal},
Multi-SWE-bench~\citep{multiswebench2025}, and
SWE-Compass~\citep{xu2025swecompass} move toward realistic issue resolution
and agentic coding. These benchmarks are important for measuring program
synthesis, bug fixing, and repository maintenance, but they do not directly
evaluate mobile app feature implementation from behavioral specifications in a
runnable platform-specific project. \sysa targets this missing setting for
OpenHarmony ArkTS applications.

\paragraph{Mobile, GUI, and app-generation benchmarks.}
Mobile-agent benchmarks such as Android in the
Wild~\citep{rawles2023aitw}, AndroidWorld~\citep{rawles2025androidworld},
SPA-Bench~\citep{chen2025spabench}, Mobile-Agent~\citep{wang2024mobileagent},
B-MoCA~\citep{lee2025bmoca}, and Mobile-Bench~\citep{deng2024mobilebench}
evaluate agents operating existing apps through device or GUI actions.
Mobile GUI datasets also support data-driven research on interface
understanding and cross-device GUI development~\citep{hu2023papt}.
MobileDev-Bench~\citep{mobiledevbench2026} brings issue resolution to Android
Native, React Native, and Flutter apps. On the generation side,
WebApp1K~\citep{cui2024webapp1k}, DesignBench~\citep{xiao2025designbench},
WebCoderBench~\citep{liu2026webcoderbench}, PlayCoder/PlayEval~\citep{peng2026playcoder},
and AppEvalPilot~\citep{appevalpilot2025} study web, front-end, GUI, or
generated-application evaluation. Mobile app quality can also depend on performance-sensitive build and binary
properties, such as compiler optimization settings in native
libraries~\citep{hu2026optdetect}.
These efforts share our interest in
interactive behavior, but most evaluate either action completion in existing
apps, issue-level patch correctness, or coarse generated-application quality.
They do not test whether an agent can modify OpenHarmony source code so that a
specified app feature becomes buildable and behaviorally correct.

\paragraph{OpenHarmony-specific evaluation.}
Work targeting the OpenHarmony ecosystem remains sparse. ArkEval~\citep{arkeval2026}
benchmarks automated repair for ArkTS, and framework-aware code generation
studies synthesize training data from API knowledge graphs to improve
OpenHarmony code generation~\citep{harmonyos_kg_codegen2025}. These efforts
operate mainly at the function, API, or repair level. In contrast,
OpenHarmony app development requires coordination across ArkUI components,
routing, state, persistence, platform APIs, command-line build tooling, device
connectors, and UI testing infrastructure~\citep{huawei_arkui_docs,huawei_hvigor_docs,openharmony_hdc_repo,huawei_uitest_docs}.
Public benchmark coverage remains limited for app-level feature implementation
in complete OpenHarmony projects.

\paragraph{Benchmark gap.}
Across these lines, existing evaluation typically reduces to final patch
correctness, outcome-level UI task completion, or broad generated-application
quality. \sysa addresses a different gap: it evaluates source-level
implementation of platform-specific app features from behavioral task inputs.
For the spec-driven source, each task is decomposed into fine-grained F-points,
and observable behavior is checked through executable OpenHarmony UI tests.
Static functional coverage and code-quality assessment are retained as
construction-time checks or optional diagnostics rather than primary leaderboard
metrics (\autoref{sec:evaluation-dimensions}).

\section{Limitations}
\label{sec:limitations}

The current report has three main limitations. First, all reported results use
DevEco Code as the fixed agent framework. This design keeps the agent scaffold,
tool access, prompt template, build scripts, and scoring pipeline fixed, so the
main comparison isolates differences among the evaluated underlying models as
much as possible. At the same time, the reported numbers should be interpreted
as DevEco Code--model configurations rather than as a direct comparison of
multiple independent coding-agent frameworks.

Second, each configuration is evaluated with three independent full-suite runs.
This repeated-run design is sufficient to report an observed run-level range
and to avoid a best-of-three score, but it does not provide an inferential
confidence interval or a statistical separation between adjacent leaderboard
entries. Fine-grained rank differences should therefore be read cautiously,
especially when run-level ranges overlap or rounded mean values are close.

Third, the bug-fix source is derived from historical public pull requests. This
choice grounds the tasks in real OpenHarmony development, but it also means
that possible pre-training exposure cannot be excluded for that source. We
therefore treat the reported numbers as a dated benchmark snapshot and
recommend recording benchmark versions, task snapshots, evaluation-script
snapshots, and model identifiers for future leaderboard entries.

\section{Conclusion}
\label{sec:conclusion}

We introduced \sysa, an app-level benchmark for evaluating LLM-based coding
agents on OpenHarmony ArkTS development. The benchmark spans three complementary
task sources: new-feature, spec-driven, and bug-fix. Together, they contain 153 top-level inputs
and 242 F-points across 42 OpenHarmony repositories. The
leaderboard is scored over top-level tasks; F-points provide the executable
verification granularity within those tasks.

\sysa evaluates each Agent--LLM configuration through a shared build-and-test
pipeline. Task Completion measures whether each top-level task builds and
satisfies all required executable checks. In the spec-driven source, this means
all F-point checks inside a task must pass. First Build Success Rate, Final
Build Success Rate, and run-level resource metrics provide complementary
views of delivery reliability and evaluation resource use. This design separates code
that merely builds from code that implements the required application behavior.

Each Agent--LLM configuration is evaluated through three independent
full-suite runs. We compute every effectiveness and build rate separately for
each run and report the arithmetic mean. Over the combined 153-task suite,
DevEco Code with GLM-5.2 achieves the highest mean Task Completion at
58.39\%, with an observed run-level range of [56.21\%, 60.78\%]. Kimi-K3 and
Qwen3.8-Max-Preview follow at 57.30\% and 55.99\%, respectively. These results
show substantial model- and task-level variation under the fixed DevEco Code
agent framework used in this report.

The leading configuration uses a mean of 2.57 million tokens and 1936.4 seconds
per task. Qwen3.8-Max-Preview uses the fewest tokens and shortest wall-clock
time in the current table, at 0.98 million tokens and 394.3 seconds per task.

The results also quantify the relationship between buildability and functional
correctness. Mean Final Build Success Rate ranges from 94.77\% to 100.00\%,
while mean Task Completion ranges from 48.36\% to 58.39\%. Successful compilation does
not by itself demonstrate correct UI behavior, state transitions, persistence,
or platform integration. Executable application-level checks are therefore
necessary for measuring whether a coding agent has completed the requested
work.

By combining realistic OpenHarmony projects, F-points, and
Agent--LLM evaluation, \sysa supports reproducible comparison of incremental
implementation, specification following, and bug fixing in the OpenHarmony
ecosystem.

\section*{Contributors}
\addcontentsline{toc}{section}{Contributors}
\phantomsection\label{sec:contributors}

\noindent
Li Li, Han Hu, Tianjian Zhang, Xin Peng, Fangzhu Mao, Qingyu Zhang, Xiaoheng Xie, Zhongmin Tang, Zhihao Lin, Haolin Ruan, Miaomiao Dong, Liuchuan Zhu, Yue Li, Chi Chen, Wenkang Zhong, Mingfei Zhang, Yang Yu, Bo Sun, Chaorui Zhang, Weixi Zhang, Wei Han, Bo Bai, Kui Liu, Gang Fan, Siru Liu, Jiaqian Zhou, Jiali Sun, Yunbiao Dong, Wenhao Zhong, Yunhong Xu.

\section*{Acknowledgments}
\addcontentsline{toc}{section}{Acknowledgments}

\noindent
We thank the Huawei Hong Kong Research Center, the Huawei Developer Platform Department, and related experts for their valuable guidance on dataset design, real-world OpenHarmony application evaluation, and trajectory analysis. In particular, we thank Ning Jiang, Jiamin Wu, Jinyu Chen, Heqing Huang, and Yu Liu for their helpful feedback and support.

\clearpage
\bibliography{references}
\bibliographystyle{harmonybench}

\end{document}